# Velocity distribution of high-energy particles and the solar neutrino problem

Jian-Miin Liu*
Department of Physics, Nanjing University
Nanjing, The People's Republic of China
*On leave. Present mailing address: P.O.Box 1486, Kingston, RI 02881, USA.

**Abstract**
   High energy infers high velocity and high velocity is a concept of special relativity. The Maxwellian velocity distribution is corrected to be consistent with special relativity. The corrected distribution reduces to the Maxwellian distribution for small velocities, contains a relatively depleted high-energy tail and vanishes at the velocity of light. This corrected distribution will lower solar neutrino fluxes and change solar neutrino energy spectra but keep solar sound speeds.

**Introduction**.   The solar neutrino problem is a long-standing puzzle in modern physics. Besides the difficulty in understanding the observed data of the $^8$B-, $^7$Be-, CNO-, pp- and pep-neutrino fluxes in different experiments from the viewpoint of solar neutrino energy spectra, it contains discrepancies between the measured solar neutrino fluxes and those predicted by standard solar models which discrepancies have existed for more than thirty years [1-14]. The measured solar neutrino fluxes range from 33%+/-5% to 58%+/-7% of the predicted values [1-4,7]. Since standard solar models lead to a very close agreement about sound speeds, better than 0.2%, between theoretical calculations and helioseimological observations [2,5], the difficulty and the discrepancies are regarded as an evidence for new physics. Something new is required for standard solar models to lower the calculated solar neutrino fluxes and to change solar neutrino energy spectra. Massive neutrinos, neutrino flavor oscillations and lepton flavor non-conservation beyond standard electroweak model have been suggested for this new physics [3,6,7]. However, this kind of new physics is still waiting to be found and confirmed out of solar neutrino arena.
   In this letter, we suggest another kind of new physics which concerns a correction to the Maxwellian velocity distribution for high-energy particles.

**A personal opinion**.       Solar core is a dense plasma of high temperature $T_c = 14.9 \times 10^6$ K and high density $\rho_c = 150 g/cm^3$. Solar core, to our knowledge, produces and radiates its neutrinos primarily through nuclear fusion reactions in the proton-proton cycle and the CNO cycle. In solar interior, a proton or nucleus must penetrate the repulsive Coulomb barrier and collides with another proton or nucleus creating a nuclear fusion reaction. On classical mechanics, as the height of Coulomb barrier is far above thermal energy $K_B T_c$: their ratio is typically greater than a thousand [2], nuclear fusion reactions can occur only in those pairs of protons or nuclei having high relative energies. From the viewpoint of quantum mechanics, as the tunnel effect lets a proton or nucleus penetrate through the repulsive Coulomb barrier of another proton or nucleus with a probability which exponentially decreases with relative energy decreasing, most nuclear fusion reactions occur in those pairs of protons or nuclei having high relative energies. On the other hand, at the temperatures and densities in solar interior, the interacting protons and nuclei reach their equilibrium distribution in such a short time that it is infinitesimal compared to the mean lifetime for a nuclear fusion reaction [2]. An equilibrium velocity distribution is applicable enough to estimating the rates of these nuclear fusion reactions. In solar interior, therefore, it is the equilibrium velocity distribution of high-energy protons and nuclei that participates in determining solar nuclear fusion reaction rates, solar neutrino fluxes and solar neutrino energy spectra.

This equilibrium velocity distribution of high-energy protons and nuclei is universally accepted as the Maxwellian velocity distribution. But, high energy infers high velocity and high velocity is a concept of special relativity. So, in our opinion, we need to look for an equilibrium velocity distribution which is consistent with Special Relativity (SR-consistent, hereafter) for these high-energy protons and nuclei, in other words, we need to correct the Maxwellian velocity distribution. Being SR-consistent, the corrected distribution must reduce to the Maxwellian distribution for small velocities, vanish at the velocity of light and be nothing for any velocity faster than the velocity of light.

Some physicists noticed and pointed out that a progressive depletion of the high-energy tail of the Maxwellian distribution can lower solar nuclear reaction rates and solar neutrino fluxes without changing solar bulk properties. They proposed some velocity distributions having a depleted high-energy tail. Clayton [8,9] suggested

$$\approx \exp\{-my^2/2K_BT - \delta(my^2/2K_BT)^2\} \tag{1}$$

as early as 1974, where parameter $\delta$ was determined to be $\delta \geq 0.01$ in comparison with experimental facts. In Eq.(1), y is the velocity magnitude, $y=(y^ry^r)^{1/2}$, r=1,2,3. Kaniadakis [10,12], Corddu [11], Gervivo [13], Lavgno [14] and others considered the Tsallis distribution,

$$\approx [1+(q-1)\frac{my^2}{2K_BT}]^{1/(1-q)} \theta[1+(q-1)\frac{my^2}{2K_BT}], \tag{2}$$

where $\theta$ is the Heaviside step function and q is a parameter to be determined. Both Clayton's distribution and the Tsallis distribution are non-vanishing for the velocity of light and velocities greater than the velocity of light. They are not SR-consistent.

**Velocity space.** Due to the sharp conflict between the concepts in statistical mechanics based on pre-relativistic mechanics and those in special relativity, so far, we have not had an acceptable Lorentz-invariant statistical mechanics, from which we can deduce an SR-consistent equilibrium velocity distribution [15]. Fortunately, we may have such an equilibrium velocity distribution through analyzing velocity space.

Velocity space is a space in which pairs of points represent relative velocities. The three-dimensional velocity space defined by

$$dY^2 = H_{rs}(y)dy^rdy^s, \quad r,s=1,2,3, \tag{3a}$$

$$H_{rs}(y) = c^2\delta^{rs}/(c^2-y^2) + c^2y^ry^s/(c^2-y^2)^2, \quad \text{real } y^r \text{ and } y<c, \tag{3b}$$

in the usual velocity-coordinates $\{y^r\}$, r=1,2,3, where $y^r$ is the well-defined Newtonian velocity, $y=(y^ry^r)^{1/2}$, and c is the velocity of light, has been studied for many years [16]. This velocity space is characterized by a finite boundary at c and the Einstein velocity addition law.

Mathematically, this velocity space can be represented in terms of the so-called primed velocity-coordinates $\{y'^r\}$, r=1,2,3, which are connected with the usual velocity-coordinates by

$$dy'^r = A^r_s(y)dy^s, \quad r,s=1,2,3, \tag{4a}$$

$$A^r_s(y) = \gamma\delta^{rs} + \gamma(\gamma-1)y^ry^s/y^2, \tag{4b}$$

where $\gamma=1/(1-y^2/c^2)^{1/2}$. The represented velocity space has the Euclidean structure,

$$dY^2 = \delta_{rs}dy'^rdy'^s, \quad r,s=1,2,3, \tag{5}$$

in the primed velocity-coordinates because

$$\delta_{rs}A^r_p(y)A^s_q(y) = H_{pq}(y), \quad r,s,p,q=1,2,3.$$

With standard calculation techniques in Riemann geometry, we can find
$H^{rs}(y) = (c^2-y^2)\delta^{rs}/c^2 - (c^2-y^2)y^ry^s/c^4$,
$\Gamma^i_{jk} = \quad 2y^i/(c^2-y^2)$, if i=j=k;
$\quad\quad\quad y^k/(c^2-y^2)$, if $i=j \neq k$;
$\quad\quad\quad y^j/(c^2-y^2)$, if $i=k \neq j$;
$\quad\quad\quad 0$, otherwise,

where $H^{rs}(y)$ is the contravariant metric tensor and $\Gamma^i_{jk}$ is the Christoffel symbols. The equation of geodesic line is therefore

$$\ddot{y}^r + [2/(c^2-y^2)]\dot{y}^r(y^s\dot{y}^s) = 0, \quad r,s=1,2,3, \tag{6}$$

where dot refers to the derivative with respect to velocity-length. Introducing new variables

$$w^r = \dot{y}^r/(c^2-y^2), \quad r=1,2,3, \tag{7}$$

we are able to rewrite Eq.(6) as

$$\dot{w}^r = 0, \quad r=1,2,3. \tag{8}$$

It is seen that

$$w^r = \text{constant}, \quad r=1,2,3, \tag{9}$$

are a solution to Eqs.(8). Due to Eqs.(7) and (8), we have

$$w^s y^r - w^r y^s = \text{constant}, \quad r,s=1,2,3. \tag{10}$$

Eqs.(9) and (10) specify three linear relations between any two of $y^1$, $y^2$ and $y^3$. These linear relations give the following shape to the equation of geodesic line between two points, $y_1^r$ and $y_2^r$, $r=1,2,3$,

$$y^r = y_1^r + \alpha(y_2^r - y_1^r), \quad 0 \leq \alpha \leq 1, \quad r=1,2,3. \tag{11}$$

Using Eqs.(11), at some length, we can find velocity-length between two points $y_1^r$ and $y_2^r$,

$$Y(y_1^r, y_2^r) = \frac{c}{2} \ln \frac{b+a}{b-a}, \tag{12a}$$

$$b = c^2 - y_1^r y_2^r, \quad r=1,2,3, \tag{12b}$$

$$a = \{(c^2 - y_1^i y_1^i)(y_2^j - y_1^j)(y_2^j - y_1^j) + [y_1^k(y_2^k - y_1^k)]^2\}^{1/2}, \quad i,j,k=1,2,3. \tag{12c}$$

In the case of $y_1^r = 0$ and $y_2^r = y^r$, the velocity-length becomes

$$Y(0,y^r) = \frac{c}{2} \ln \frac{c+y}{c-y} \quad \text{or} \quad Y^2(0,y^r) = [\frac{c}{2y} \ln \frac{c+y}{c-y}]^2 \delta_{rs} y^r y^s, \quad r,s=1,2,3. \tag{13}$$

On the other hand, we know from Eqs.(5) that square of velocity-length between two points $y_1'^r$ and $y_2'^r$ is

$$Y^2(y_1'^r, y_2'^r) = \delta_{rs}(y_2'^r - y_1'^r)(y_2'^s - y_1'^s), \quad r,s=1,2,3, \tag{14}$$

and

$$Y^2(0, y'^r) = \delta_{rs} y'^r y'^s, \quad r,s=1,2,3, \tag{15}$$

if $y_1'^r = 0$ and $y_2'^r = y'^r$. Eqs.(13) and (15) imply

$$y'^r = [\frac{c}{2y} \ln \frac{c+y}{c-y}] y^r, \quad r=1,2,3, \tag{16}$$

$$y' = \frac{c}{2} \ln \frac{c+y}{c-y} \tag{17}$$

when $(y'^1, y'^2, y'^3)$ and $(y^1, y^2, y^3)$ represent the same point in the velocity space, where $y' = (y'^r y'^r)^{1/2}$, $r=1,2,3$. We call $y'^r$, $r=1,2,3$, the primed velocity [17,18]. Its definition from the measurement point of view is given in Ref.[17].

The Galilean addition law of primed velocities links up with the Einstein addition law of corresponding Newtonian velocities [17,18].

**Equilibrium velocity distribution.** The Euclidean structure of the velocity space in the primed velocity-coordinates $\{y'^r\}$ convinces us that the Maxwellian velocity and velocity rate distribution formulas are valid in the primed velocity-coordinates, namely

$$P(y'^1, y'^2, y'^3) dy'^1 dy'^2 dy'^3 = N(\frac{m}{2\pi K_B T})^{3/2} \exp[-\frac{m}{2K_B T}(y')^2] dy'^1 dy'^2 dy'^3 \tag{18}$$

and

$$P(y') dy' = 4\pi N(\frac{m}{2\pi K_B T})^{3/2} (y')^2 \exp[-\frac{m}{2K_B T}(y')^2] dy', \tag{19}$$

where N is the number of particles, m their rest mass, T the temperature, and $K_B$ the Boltzmann constant.

We can employ Eqs.(4a-4b) and (17) to represent these two formulas in the usual velocity-coordinates $\{y^r\}$, $r=1,2,3$. Using Eq.(17) and

$$dy'^1 dy'^2 dy'^3 = \gamma^4 dy^1 dy^2 dy^3$$

which is inferred from Eqs.(4a-4b), we have from Eq.(18),

$$P(y^1,y^2,y^3)dy^1dy^2dy^3 = N \frac{(m/2\pi K_B T)^{3/2}}{(1-y^2/c^2)^2} \exp[-\frac{mc^2}{8K_B T}(\ln\frac{c+y}{c-y})^2]dy^1dy^2dy^3. \qquad (20)$$

Using Eq.(17) and
$$dy' = \gamma^2 dy$$
which comes from differentiating Eq.(17), we have from Eq.(19),

$$P(y)dy = \pi c^2 N \frac{(m/2\pi K_B T)^{3/2}}{(1-y^2/c^2)} (\ln\frac{c+y}{c-y})^2 \exp[-\frac{mc^2}{8K_B T}(\ln\frac{c+y}{c-y})^2]dy. \qquad (21)$$

For small velocities, velocity distribution function $P(y^1,y^2,y^3)$ in Eq.(20) and velocity rate distribution function $P(y)$ in Eq.(21) respectively reduce to

$$N(\frac{m}{2\pi K_B T})^{3/2} \exp[-\frac{m}{2K_B T}(y^2)]$$

and

$$4\pi N(\frac{m}{2\pi K_B T})^{3/2} (y^2) \exp[-\frac{m}{2K_B T}(y^2)],$$

which are just the Maxwellian velocity and velocity rate distribution functions. $P(y^1,y^2,y^3)$ and $P(y)$ become nothing when $y$ is greater than $c$. They vanish as $y$ approaches $c$. Actually,

$$\lim_{y \to c} P(y^1,y^2,y^3) = \lim_{z \to +\infty} \frac{c^2 N}{4} (\frac{m}{2\pi K_B T})^{3/2} z^2 \exp\{-A[\ln(2cz)]^2\},$$

$$\lim_{y \to c} P(y) = \lim_{z \to +\infty} \frac{\pi c^3 N}{2} (\frac{m}{2\pi K_B T})^{3/2} z [\ln(2cz)]^2 \exp\{-A[\ln(2cz)]^2\},$$

where $z=1/(c-y)$ and $A = \frac{mc^2}{8K_B T}$. Since $\ln(2cz)$ is smaller than $2cz$ for large $z$, both $\lim_{y \to c} P(y^1,y^2,y^3)$ and $\lim_{y \to c} P(y)$ are smaller than $\lim_{z \to +\infty} (\text{constant})z^3\exp\{-A[\ln(2cz)]^2\}$. The last limit equals zero.
Eqs.(20) and (21) are the SR-consistent equilibrium velocity and velocity rate distributions.

The positive and monotonically decreasing high-energy tail of velocity distribution function $P(y^1,y^2,y^3)$ goes to zero as $y$ approaches $c$, while the positive and monotonically decreasing high-energy tail of the Maxwellian velocity distribution function goes to zero as $y$ approaches infinity. That indicates a depleted high-energy tail of velocity distribution function $P(y^1,y^2,y^3)$ with respect to the Maxwellian velocity distribution function. The same situation exists between velocity rate distribution function $P(y)$ and the Maxwellian velocity rate distribution function.

**A possible solution to the solar neutrino problem.** Evidently, the nuclear fusion reaction rate based on the SR-consistent equilibrium velocity distribution has a reduction factor with respect to that based on the Maxwellian velocity distribution [19]:

$$R = \frac{\tanh Q}{Q} R_M, \quad Q = (2\pi z_1 z_2 \frac{K_B T}{\mu c^2} \frac{e^2}{\hbar c})^{1/3}, \qquad (22)$$

where $R_M$ is the nuclear fusion reaction rate based on the Maxwellian velocity distribution. Since $0 < Q < \infty$, the reduction factor satisfies $0 < \tanh Q/Q < 1$. That gives $0 < R < R_M$. The reduction factor depends on the temperature T, reduced mass $\mu$, and atomic numbers $z_1$ and $z_2$ of the studied nuclear fusion reactions.

The SR-consistent equilibrium velocity distribution differs from the Maxwellian velocity distribution substantially in the part of high velocity. So, we have to substitute the SR-consistent equilibrium velocity distribution for the Maxwellian velocity distribution in those statistical calculations which are merely or mainly concerned with the high-velocity part of velocity distribution of the relevant particles or those statistical calculations where most relevant particles crowd in the high-velocity part. The calculations of solar nuclear fusion reaction rates, solar neutrino fluxes and solar neutrino energy spectra belong here. When most relevant particles crowd in the low-velocity part and this low-velocity part is involved in statistical calculations, the substitution for the Maxwellian velocity distribution is not so important. One of such calculations is about the sound speeds in the Sun, because most ions and particles crowd in the low-velocity part even at temperatures in solar whether interior or surface. The SR-consistent equilibrium velocity distribution, if adopted in standard solar models, will lower solar neutrino fluxes and change solar neutrino energy spectra but keep solar sound speeds.

ACKNOWLEDGMENT

The author greatly appreciates the teachings of Prof. Wo-Te Shen. The author thanks Prof. Gerhard Muller and Dr. P. Rucker for suggestions.